\documentclass[aps,12pt,superscriptaddress,amsfonts,amssymb,amsmath]{revtex4}
\pdfoutput=1

\usepackage{graphicx}
\usepackage{epsfig}
\usepackage{makeidx}
\usepackage{epstopdf}
\usepackage{natbib}
\usepackage{xcolor}

\begin{document}

\title{The configuration model for Barabasi-Albert networks}

\author{M.L.\ Bertotti \footnote{Email address: marialetizia.bertotti@unibz.it}}
\author{G.\ Modanese \footnote{Email address: giovanni.modanese@unibz.it}}
\affiliation{Free University of Bozen-Bolzano \\ Faculty of Science and Technology \\ I-39100 Bolzano, Italy}

\linespread{0.9}

\begin{abstract}

We develop and test a rewiring method (originally proposed by Newman) which allows to build random networks having pre-assigned degree distribution and two-point correlations. For the case of scale-free degree distributions, we discretize the tail of the distribution according to the general prescription by Dorogovtsev and Mendes. The application of this method to Barabasi-Albert (BA) networks is possible thanks to recent analytical results on their correlations, and allows to compare the ensemble of random networks generated in the configuration model with that of \textquotedblleft real\textquotedblright networks obtained from preferential attachment. For $\beta\ge 2$ ($\beta$ is the number of parent nodes in the preferential attachment scheme) the networks obtained with the configuration model are completely connected (giant component equal to 100\%). In both generation schemes a clear disassortativity of the small degree nodes is demonstrated from the computation of the function $k_{nn}$. We also develop an efficient rewiring method which produces tunable variations of the assortativity coefficient $r$, and we use it to obtain maximally disassortative networks having the same degree distribution of BA networks with given $\beta$. Possible  applications of this method concern assortative social networks.

\end{abstract}

\maketitle

\section{Introduction}
\label{introduction}

In spite of the large number of existing studies on Barabasi-Albert (BA) networks, their two-point correlation functions have been completely analysed only recently by Fotouhi and Rabbat \cite{fotouhi2013degree}, who have given the full expressions of the conditional probabilities $P(h|k)$ in the large network limit for any value of the parameter $\beta$ (the number of 
{child} nodes in the preferential attachment process). 

Concerning the assortativity properties of BA networks, in previous work some estimates of the Newman coefficient $r$ were found \cite{newman2002assortative}. According to these estimates, for large $N$ (number of nodes), $r$ vanishes as $-\ln^2 N/N$. It was therefore generally believed that BA networks are almost uncorrelated, and numerical simulations appeared to confirm this. However, more recent asymptotic estimates \cite{fotouhi2018temporal,bertotti2018bass} yield a different result: $r$ vanishes only as $-\ln^2 N/\sqrt{N}$ for large $N$. It should be recalled that for real networks with the same scale-free exponent ($\gamma=3$), the $r$ coefficient is always small in absolute value, so even this small total disassortativity is significant.

By computing the function $k_{nn}(k)$ of BA networks (average nearest neighbor degree of a node of degree $k$) we have shown in \cite{bertotti2018bass} that it is strongly decreasing for small $k$ and slowly increasing for large $k$. This means that the total slight disassortativity measured by the $r$ coefficient is in fact the result of an unexpected mixed assortative/disassortative behavior of these networks.

This peculiar structural property may have an influence on the dynamics on BA networks. In particular, we have shown in  \cite{bertotti2018bass} for the Bass innovation diffusion model on a network, that finite BA networks exhibit the fastest diffusion among scale-free networks with exponent $\gamma=3$. This result was obtained by solving numerically the coupled nonlinear differential mean-field equations of the model \cite{bertotti2016bass,vespignani2012modelling} (a system of $n$ equations for a network with maximum degree $n$) and finding the time of the diffusion peak {in dependence on the network structure}. 
Comparisons were made between BA networks, uncorrelated networks, disassortative networks built according to a method by Newman \cite{newman2003mixing} and assortative networks built with a recipe we have recently developed \cite{bertotti2016bass}. Results are displayed in Tab.\ \ref{table1}. The networks employed had typically maximum degree $n=10^2$, which corresponds to $N\approx 10^4$ nodes. It turns out that the fastest diffusion process occurs on BA networks with $\beta=1$. Note that the peak diffusion time in the Bass model is, unlike in the SI model, independent from the initial conditions, and gives therefore useful information on the diffusion features of the network.

\begin{table}
\begin{center}
\begin{tabular}{|@{}l|ccc@{}|} 
\toprule
\quad \  \ & \  \ & {\bf{Diffusion times}} \ & \ \   \nonumber \\
\hline
\ & \  \ & \ \  \ & \ \  \  \nonumber \\
\ \ network  \ \ & \ \ $q = 0.30$ \ &\ $q = 0.38$ \ & \ $q = 0.48$ \ \  \nonumber \\
\ & \  \ & \ \  \ & \ \  \  \nonumber \\
\hline
\quad \ BA1 \ & \ 4.84  \ & \ 4.22  \ & \ 3.64     \nonumber \\
\quad \ BA2 \ & \ 5.20  \ & \ 4.54  \ & \ 3.94   \nonumber \\
\quad \ BA3 \ & \ 5.40  \ & \ 4.74  \ & \ 4.10      \nonumber \\
\quad \ UNC \ & \ 5.50  \ & \ 4.84   \ & \ 4.20        \nonumber \\
\quad \ BA4 \ & \ 5.54  \ & \ 4.86  \ & \ 4.22     \nonumber \\
\quad \ BA5 \ & \ 5.66  \ & \ 4.96  \ & \ 4.32   \nonumber \\
\quad \ DIS \ & \ 5.76  \ & \ 5.06  \ & \ 4.40      \nonumber \\
\quad \ ASS \ & \ 6.00  \ & \ 5.50  \ & \ 4.98     \nonumber \\
\hline
\botrule
\end{tabular}	
\caption{Times (in years) of the diffusion peak in the Bass diffusion model on networks for eight different scale-free networks with exponent $\gamma=3$, 
for three different values of the imitation coefficient $q$.  All networks have maximum degree $n=100$ ($N\simeq 10^4$). 
Those here considered are BA networks with values of $\beta$ equal to $1, 2, 3, 4, 5$ (resp. denoted BA1, BA2, BA3, BA4, BA5),
uncorrelated networks (UNC), disassortative and assortative networks (resp., DIS and ASS).
The values of the $q$ coefficient 
here considered,
$0.30$, $0.38$, $0.48$, 
belong to a range of typical realistic values in innovation diffusion theory \cite{jiang2006virtual} 
and the publicity coefficient is set to the value $p=0.03$, which is another realistic value \cite{jiang2006virtual}.
The assortativity coefficients of the BA networks are respectively $r=-0.104,-0.089, -0.078,-0.071,-0.065$. The disassortative network is built with the Newman recipe  \cite{newman2003mixing,bertotti2018bass}
with $d=4$  and has $r=-0.084$. The assortative network is built with our recipe  \cite{bertotti2016bass,bertotti2018bass} with $\alpha=1/2$, and has $r=0.863$.}		
\label{table1}
\end{center}
\end{table}

In this work we use for the first time the correlation functions found in \cite{fotouhi2013degree} in order to build in the configuration model networks which display these peculiar correlations, and investigate their properties.

The configuration model \cite{newman2010networks} is a method for the generation of random networks having an assigned degree distribution. It is therefore a powerful extension of the original concept of random network introduced by Erd\"os, and has been extensively studied with analytical and numerical methods, especially for the case of scale-free networks. Classical results concern the conditions for the formation of a giant component \cite{molloy1995critical} and its clustering features. 

Some authors have also raised the question of whether it is possible to generate networks with pre-assigned correlations. Newman has proposed for this purpose in \cite{newman2003mixing} a method based on a degree-preserving rewiring procedure; more recently, the issue has been also discussed by Bassler et al.\ \cite{bassler2015exact} and by Boguna et al.\ \cite{boguna2003class}. 
The practical applications of these ideas have been, until now, rather limited. Yet, from the applicative point of view the possibility of an efficient generation of networks with given correlations is quite attractive. 

For example, social networks are known to be generally assortative, and in order to study diffusion processes on these networks in the mean-field approximation it is very useful to construct mathematically certain families of assortative correlation matrices \cite{bertotti2016bass}. If it is possible to produce explicit realizations of networks with such correlations, these can be used to obtain a further characterization of the diffusion process, possibly also with agent-based methods etc. In fact, an assortative rewiring has been proposed already in \cite{newman2003mixing} and in \cite{xulvi2004reshuffling}, but with some limitations; in the first case the assortative matrices employed do not generally satisfy a positivity criterium, in the second case no correlations matrices are employed, and the rewiring criterium works on an heuristic basis.

With the above applications in mind, our aim in this work is to use the correlations matrices of BA networks and the rewiring procedure by Newman to test under controlled conditions the configuration model with pre-assigned correlations. In fact, one of the features of BA networks which makes them so popular and widely used for the simulation of real networks is that they can be readily generated via a preferential attachment procedure. Since their correlation matrices are now available, by re-constructing them in the configuration model we can compare the features of the resulting ensemble of networks with those of the networks produced by preferential attachment. As we shall see, this gives useful insights on the method in general.

The rest of the paper is organized as follows.
In Section\ \ref{conf} we discuss the mentioned re-wiring procedure which generates by using the configuration model
an ensemble of networks having as prescribed correlations the correlations of BA networks.
Some features of the networks of the ensemble obtained in this way are then discussed, including the behavior of their average nearest neighbours degree function $k_{nn}(k)$.
The re-wiring procedure is also adapted in Section\ \ref{max} to generate maximally disassortative scale-free networks having the same exponent as the BA ones. 
Section\ \ref{conclusion} concludes by discussing the results and some potential follow-ups.
Finally, in the Appendix some definitions and the expressions for the case of BA networks of some quantities
used throughout the paper are recalled.

\section{The configuration model with Newman rewiring}
\label{conf}

\subsection{Discretized degree distribution}
\label{ddd}

Given the total number $N$ of nodes in the network, suppose that we want to assign to each node a degree $D_1, \ldots , D_N$ according to a degree distribution with the form of a power law $P(k) =c_\gamma/k^\gamma$. As discussed in \cite{boguna2004cut} the maximum node degree $n$ present in the network can be obtained from $N$ through the relation
\begin{equation}
\int_n^\infty P(k) dk = \frac{1}{N} \, .
\label{int-cri}
\end{equation}
This means that $n$ is the degree above which one expects to find at most one node.
For the case of $\gamma=3$, we obtain $n \simeq \sqrt{N}$.

In practice one can set, for finite networks, $P(k)=0$ for $k>n$, and normalize $P(k)$ accordingly, by defining
\begin{equation}
c_\gamma^{-1}=\sum_{k=1}^n P(k) \, .
\end{equation}
Then one can define
\begin{equation}
N_k = {\rm Round} [P(k)N]
\label{rou}
\end{equation}
as the average number of nodes with degree $k$ present in the network, where ``Round'' denotes rounding to the nearest integer.

In this way, however, we find that $N_k$ becomes zero when $k>n_1\simeq (2Nc_\gamma)^{1/\gamma}$, which is considerably smaller than the value $n$ given by the integral criterium (\ref{int-cri}). The reason is that we are essentially discarding the fractional expectation values found from (\ref{rou}), instead of cumulating them as in (\ref{int-cri}).

This procedure has been employed in the influential paper \cite{aiello2000random} in order to generate scale-free networks with the configuration model. This work, however, pre-dates Ref.\ \cite{boguna2004cut} and the widespread use of preferential attachment for the generation of scale-free networks, especially of the BA type. Actually it is immediate to realize, by plotting the degree distribution of finite BA networks generated via preferential attachment, that a random succession of hubs in the degree interval $k \in [n_1,+\infty]$ is always present. These hubs play an important role in several dynamical processes on the network.

Therefore we shall use in the following, to obtain the discretized degree distribution $N_k$, not the simple recipe (\ref{rou}) but {one of
three different improvements of it, which give practically equivalent results for the networks considered in this work:}

(1) \emph{\textquotedblleft Cumulation\textquotedblright \, method}. In this method, for $k>n_1$ the values of $P(k)N$ are cumulated, as $k$ increases, until their sum exceeds 1; at this point, one hub is created, the cumulation procedure starts again, and so on. 

(2) \emph{\textquotedblleft Random hubs\textquotedblright \, method}. The idea is the following: if the average number of nodes with degree $k$ is smaller than 1, say $NP(k)=X<1$, then a node with this degree will be created in each realization with probability $X$. Extending the procedure to all degrees, a random variable $\xi \in (0,1)$ is generated for each value of $k$, and then denoting by Int($NP(k)$) the integer part of $NP(k)$ and by  Dec($NP(k)$) its decimal part, one sets $N_k={\rm Int}(NP(k))$ if $\xi>{\rm Dec}(NP(k))$ and $N_k={\rm Int}(NP(k))+1$ if $\xi<{\rm Dec}(NP(k))$. The number of nodes is therefore not fixed, with random variations 
of $1$ for each degree (in particular, with values $0$ or $1$ in the tail of the distribution), such to respect the degree distribution in an ensemble average.

{(3) The most general way for generating the degrees of the nodes is to use a \textit{probability transformation method}. For this one needs to define first a vector $F_k=\sum_{j=1}^k P(j)$, $F(0)=0$, where $k=1,\ldots,n$ and $P(j)$ denotes the normalized degree distribution. The values of $F_k$ define breakpoints of the unit interval (0,1). After generating a random number $\xi$ in this interval, a new node is introduced with degree $k$ if $F_{k-1}<\xi<F_k$, and the procedure is repeated $N$ times. This method has the advantage of allowing the generation of exactly $N$ nodes.}

\subsection{Description of the wiring and re-wiring algorithm}
\label{wir}

After a degree $D_i$ has been assigned to each candidate node (or ``stub'') $i$, in the classical configuration model a certain number of links is randomly attached to the stubs, until each stub reaches its planned degree. In our algorithm this wiring procedure is not random, but partially deterministic. This is more efficient and makes sense because the wiring is followed by a massive random re-wiring phase (see below) which preserves the degrees of the nodes but makes the correlations close, in an ensemble average, to the ``target'' correlations $e^0_{jk}$.

The wiring procedure starts from Node 1, whose degree $D_1$ is the largest in the network. Among the remaining ones, $D_1$ distinct nodes are chosen randomly and connected to it. For each of the nodes chosen, the number of available stubs is decreased by one. Then the same steps are repeated for Node 2 and so on, with exclusion of nodes whose stubs are all already connected. The final product will be a list of $L$ links of the form $(a,b)$, where $a$ and $b$ denote two nodes ($a,b=1,\ldots,N$). Provided $N$ is even, we have 
\begin{align}
	L=\frac{1}{2} \sum_{j=1}^N D_j= \frac{1}{2} \sum_{k=1}^n kN_k = \frac{1}{2} \sum_{k=1}^n kP(k)N=\frac{1}{2} N \left\langle k \right\rangle  \, .
\end{align}
         
For the rewiring according to the Newman procedure, we choose at random in the list of the links two links $(a,b)$ and $(c,d)$. Denote with $A,B,C,D$ the excess degrees of these nodes and define the quantities
\begin{align}
	E_1=e^0_{AB}e^0_{CD} \, , \qquad E_2=e^0_{AC}e^0_{BD} \, ,
\end{align}
where $e^0_{jk}$ is the ``target'' correlation matrix that we want to approach in the rewiring. Then
\begin{itemize}
	\item If $E_1=0$ the rewiring is performed, i.e., the links  $(a,b)$, $(c,d)$ are replaced by $(a,c)$, $(b,d)$.
	\item If $E_1>0$, we define $P=E_2/E_1$ and then generate a random number $\xi \in (0,1)$.
	\item If $P\geq 1$, the rewiring is performed.
	\item If $P<1$ and $\xi <P$, the rewiring is performed.
\end{itemize}
  
	Then another couple of links is chosen and the same steps are repeated. 
	
	The ergodicity property of this rewiring procedure has been discussed in \cite{newman2003mixing}. As empirical criterium for the attainment of equilibrium we set an average of $10^3$ rewirings per node. 
The fraction of successful rewirings for the present case of BA networks turns out to be larger than 0.5. Therefore, $10^7$ can be taken with a safe margin as a total number of attempts necessary 
for our trial networks with $N=2500$.
This can be accomplished in less than 1 second on a normal machine. The time scales linearly with the size of the network. We chose to report here on the size $N=2500$ also for practical reasons of visualization of the network and of its function $k_{nn}(k)$ (see below, Sect.\ \ref{knn}). 

{The $N$ parameter and the number of rewirings given above are only one of many possible safe choices and do not substantially affect the properties observed in the networks. Concerning the choice of the rewiring algorithm itself, we are not aware of any alternative to the Newman algorithm, if the purpose of the rewiring is to obtain networks having (in a statistical sense) pre-defined two-point ``target'' correlations.}

\subsection{Properties of BA networks obtained in the configuration model}
\label{prop}

For a BA network with $\beta=1$ (in the following also denoted as BA1), the correlation $P(1|1)$ is zero, according to the general formulas of Fotohui and Rabbat. This particular case is also obvious if one considers the totally-connected growth process of the network as obtained in the preferential attachment scheme: no node of degree 1 can be connected to a node of the same degree, otherwise an isolated pair would be formed.

\begin{figure}[t]
\begin{center}
\includegraphics[width=16cm,height=15cm]{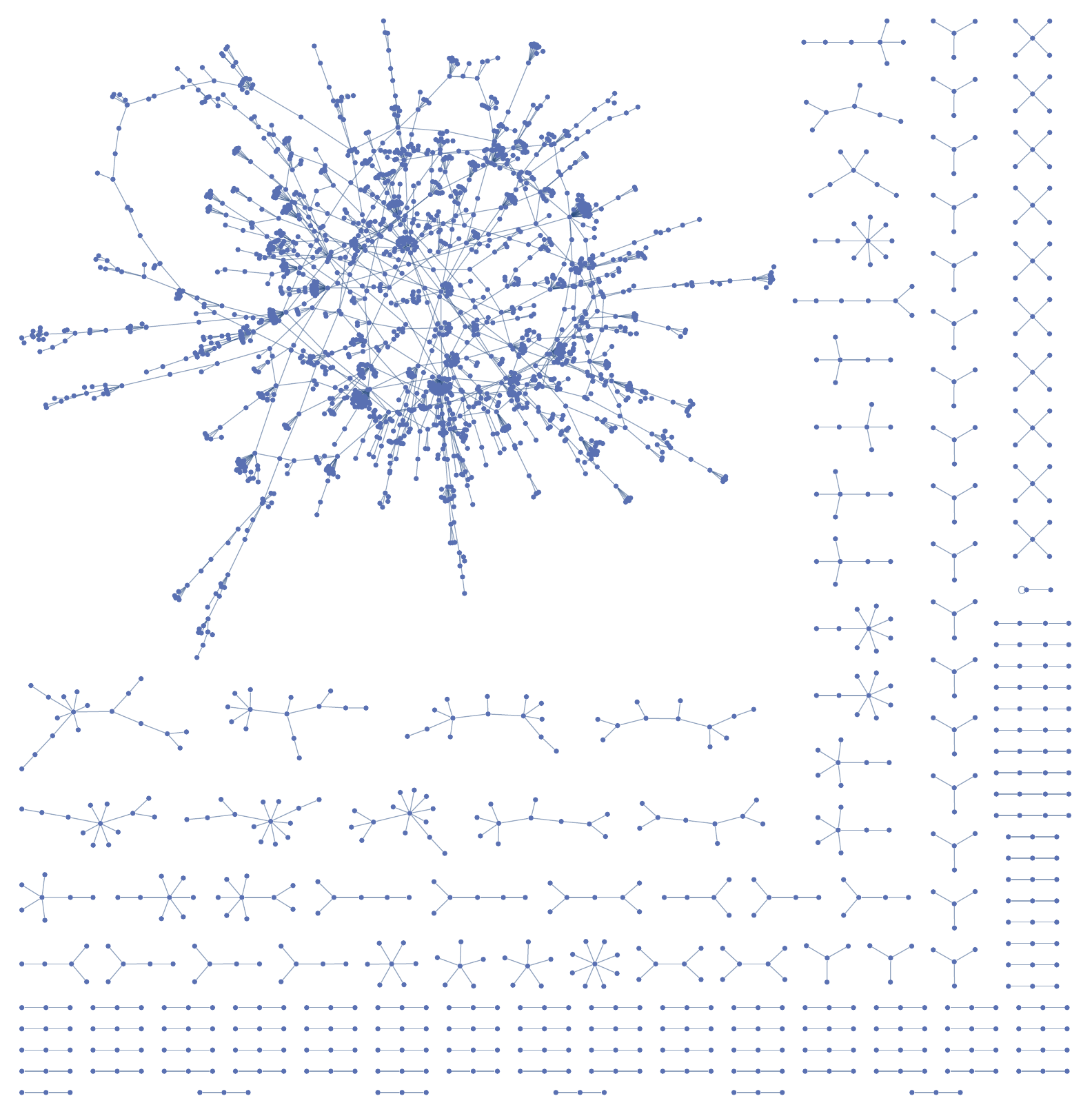}
\caption{Example of a network with $N=2500$ nodes obtained in the configuration model with Newman rewiring using the degree distribution and two-point correlations of a Barabasi-Albert network with $\beta=1$ (``BA1''). The largest hubs have degree $n\simeq 80$. The giant connected component is approx. 70\% of the total network. The small components do not include pairs (because they are forbidden by the two-point correlations), but do include triples and larger trees. The clustering coefficient is very small ($<10^{-3}$; it is exactly zero for a real BA1). One also notices in the giant component some ``small-world'' bridges which are absent in a real BA1.
} 
\label{fig1}
\end{center}  
\end{figure}

In the configuration model applied to scale-free networks with {\it random} rewiring, isolated pairs do usually form, and in large numbers, with the effect of a considerable reduction of the giant component. 
However, when we apply the configuration model to the degree distribution of a BA1 network, followed by a Newman rewiring with target correlations $e^0_{jk}$ of the BA1 type (obtained from the $P(h|k)$ as in 
eq.\ (\ref{PhkBA}) in the Appendix,
the resulting number of isolated pairs is always zero, because the condition $P(1|1)=0$ is enforced in an effective way. The size of the giant (connected) component is about $0.69\pm 0.01$. Most of the disconnected small components are triples (Fig.\ \ref{fig1}), whose origin is quite interesting.
The correlation $P(2|1)$ is non zero for BA1 networks. In fact, in the growth process with preferential attachment, connected tails of variable length can arise, in which the last node contributes to the correlation $P(2|1)$ and the intermediate nodes contribute to the correlation $P(2|2)$. When the network is re-constructed in the configuration model, isolated triples arise, because the non-vanishing conditional probability $P(2|1)$ allows to attach two nodes of degree 1 to a central node of degree 2 ``without knowing'' that on the other side of this central node there is no connection to the giant component. This is a simple demonstration of the general fact that the knowledge of the degree distribution and two-point correlations is insufficient to completely characterize a network.

\begin{figure}[t]
\begin{center}
\includegraphics[width=16cm,height=14cm]{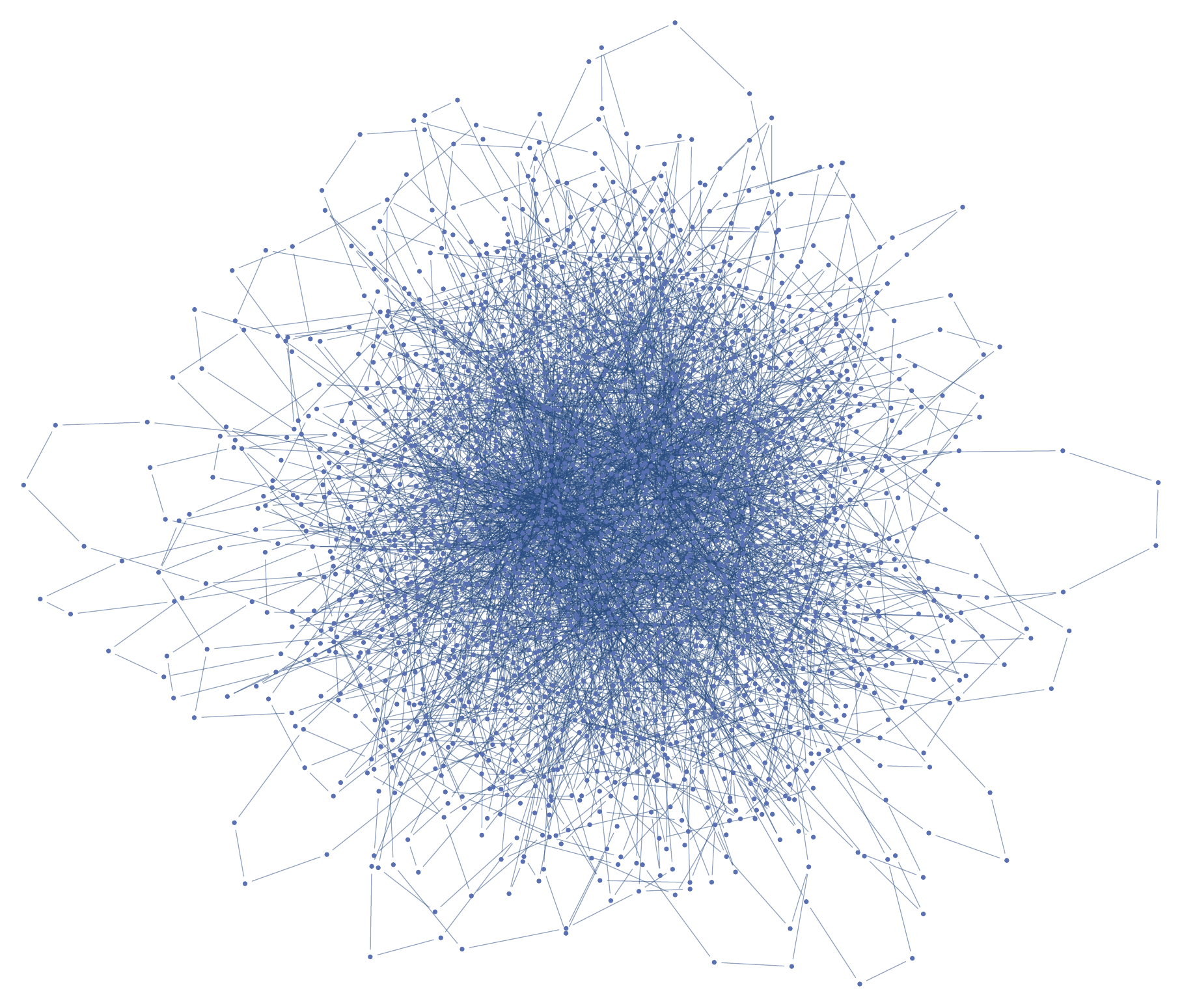}
\caption{Example of a network with $N=2500$ nodes obtained in the configuration model with Newman rewiring using the degree distribution and two-point correlations of a Barabasi-Albert network with $\beta=2$ (``BA2''). The largest hubs have degree $n\simeq 80$. The giant connected component is 100\% of the total network. This is mainly due to the large average connectivity ($\langle k \rangle \simeq 3.8$, compared to $\langle k \rangle \simeq 1.9$ of the BA1).
} 
\label{fig2}
\end{center}  
\end{figure}

For BA degree distributions with $\beta>1$ (Fig.\ \ref{fig2}) the Newman rewiring always generates a fully connected network (giant component equal to 100\%). This has little to do with the correlations, but is due instead to the large average connectivity of these degree distributions, namely $\langle k \rangle=2\beta$. For uncorrelated networks, it is known that the size of the giant component in the configuration model grows quickly as $\langle k \rangle$ grows \cite{newman2010networks}. We can readily check this with our algorithm, for example, through an initial wiring with the BA2 degree distribution, followed by a Newman rewiring with target correlations $e^0_{jk}=q_jq_k$.

\subsection{Function $k_{nn}(k)$ of BA networks obtained with the configuration model
}
\label{knn}

A possible way to check if the two-point correlation functions of the BA networks have been correctly reproduced in the configuration model is to plot the function $k_{nn}(k)=\sum_{h=1}^n hP(h|k)$, also known as average nearest neighbours degree distribution. Due to the partial summation in its definition, this function depends only on one argument and is therefore easier to analyse than the full $P(h|k)$; moreover, it has a direct qualitative interpretation in terms of assortativity and disassortativity of the network. For an uncorrelated network it is constant and equal to $\langle k^2 \rangle / \langle k \rangle$. By computing the $k_{nn}(k)$ function of the BA correlations given by Fotouhi and Rabbat, we have shown that it is decreasing at small $k$, reaching a minimum for a $k_{min}$ almost proportional to $n$ ($k_{min}\simeq 0.2 n + 20$ in the range $50\le n \le 500$), ad then it is slightly increasing for large $k$. 

\begin{figure}
\begin{center}
\includegraphics[width=8cm,height=5.5cm]{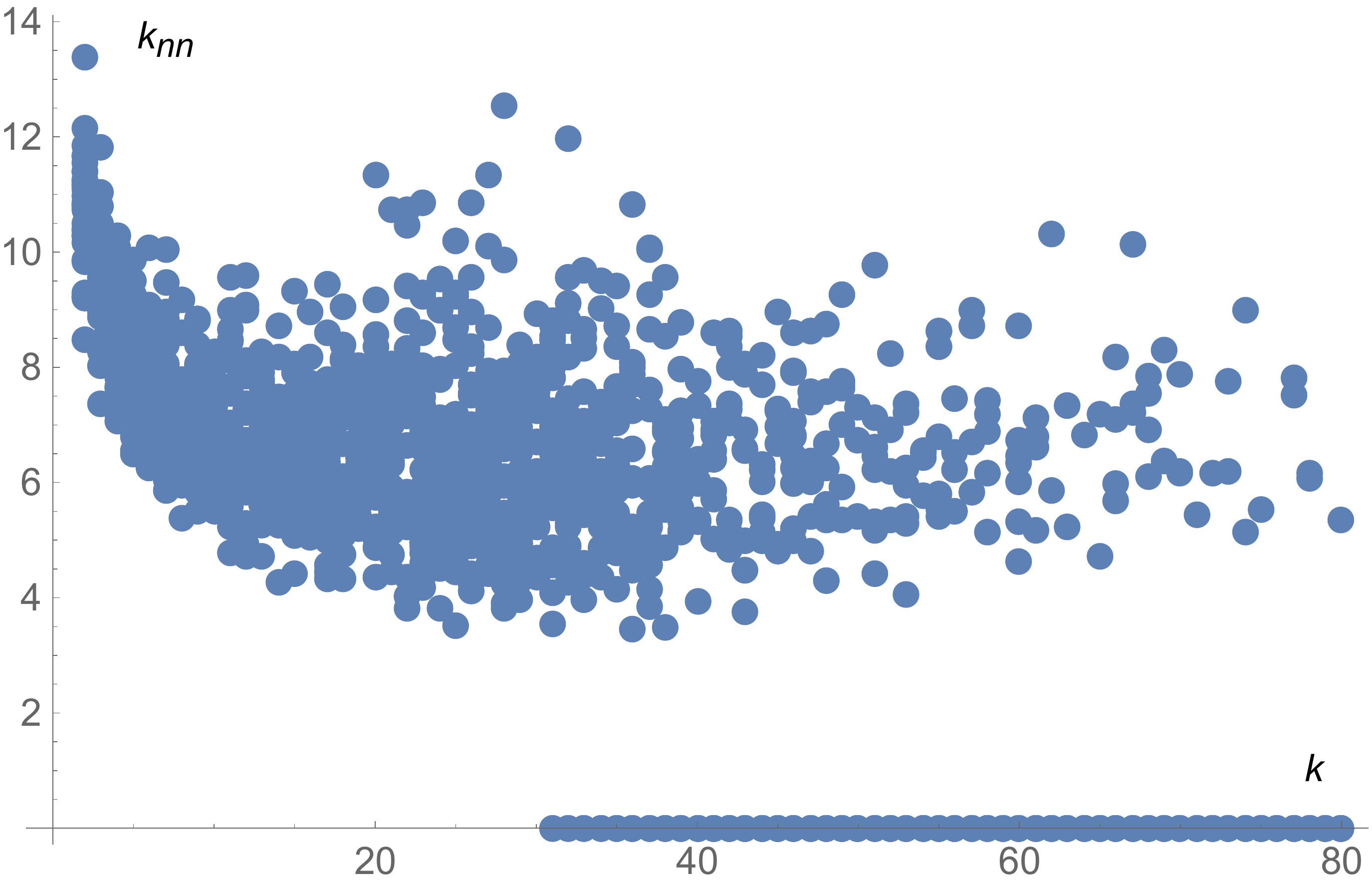}
\caption{``Cloud'' diagram of the $k_{nn}(k)$ functions of an ensemble of 20 networks generated from the configuration model with Newman rewiring using the BA2 degree distribution and correlations. The disassortativity for small $k$ is clearly visible. The hubs are obtained with the ``random hubs'' method of Sect.\ \ref{ddd}.
} 
\label{fig-cloud}
\end{center}  
\end{figure}

\begin{figure}
\begin{center}
\includegraphics[width=8cm,height=5.5cm]{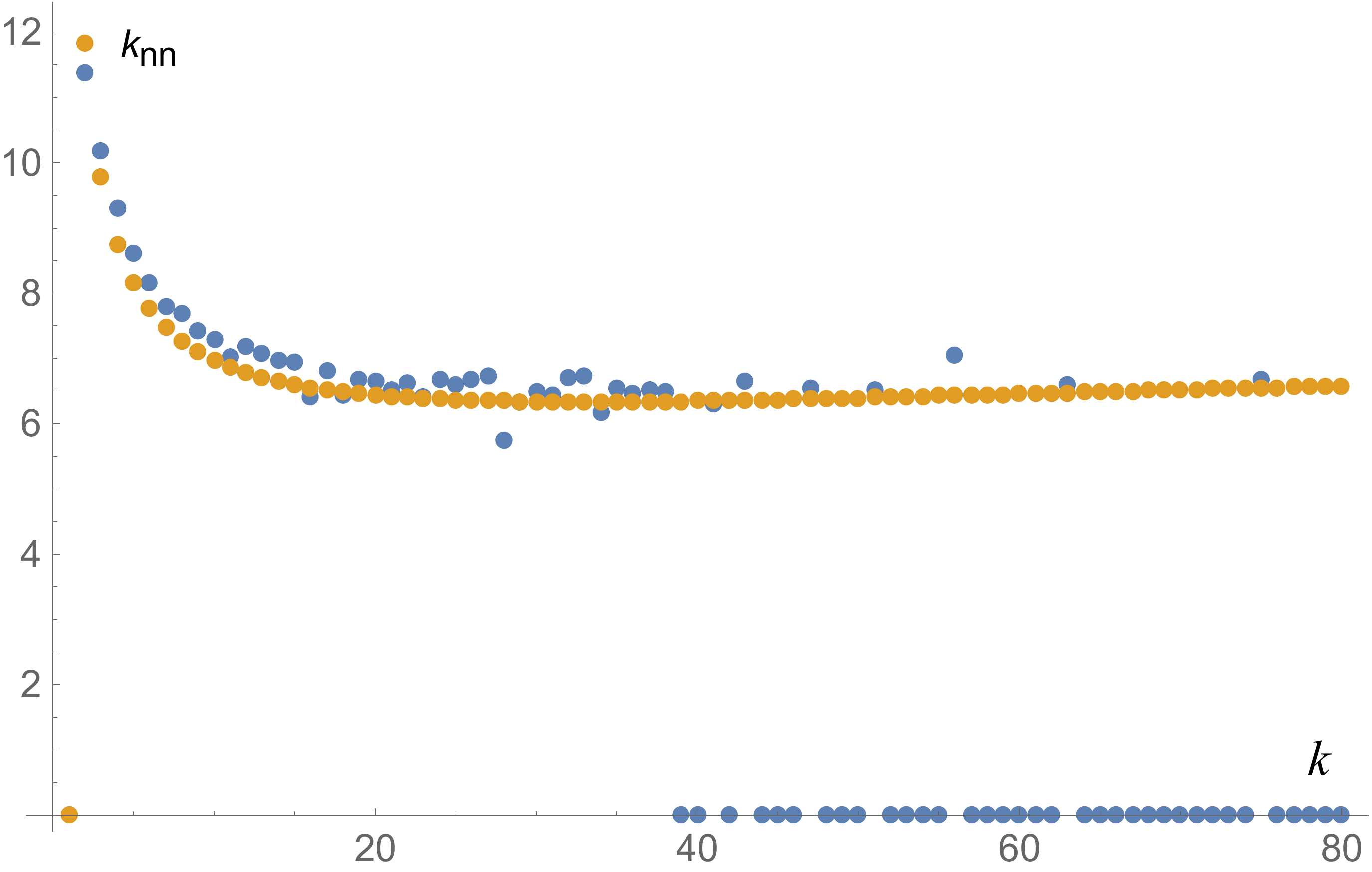}
\caption{Light point series: function $k_{nn}(k)$ for a finite BA2 network according to the formulas of Fotouhi-Rabbat. Dark point series: average of $k_{nn}(k)$ over 40 configuration model realizations obtained as in Fig.\ \ref{fig-cloud} (but with fixed hubs). The disassortativity for small $k$ is clearly visible, and the slight assortativity for large $k$ is also confirmed.
} 
\label{fig-knn-media}
\end{center}  
\end{figure}

This behavior is reproduced in the configuration model but, as expected, with large fluctuations, if one plots $k_{nn}$ for different BA networks belonging to the {\it statistical ensemble} obtained with the Newman rewiring. Graphically one can visualize such an ensemble with a ``cloud'' plot (see Fig.\ \ref{fig-cloud}). This plot is completely analogous to a cloud plot of the $k_{nn}$ functions for BA networks obtained at random with the preferential attachment algorithm. 

Note that the degrees of the largest hubs in Fig.\ \ref{fig-cloud} are randomly generated according to the ``random hubs'' method described in Sect.\ \ref{ddd}. This is very similar to what one obtains using a standard randomized preferential attachment algorithm: if one analyzes a relatively small number of realizations, one will find that the largest hubs present have variable degrees, and most of the degrees in the tail of the degree distribution are actually missing.

Alternatively to the cloud diagram, one can compute the \emph{average} of the $k_{nn}$ function over a rewiring ensemble, and compare it to the $k^{FR}_{nn}$ obtained with the $P(h|k)$ correlations of Fotouhi and Rabbat. In this case, the degrees of the largest hubs need to be fixed, otherwise the average for large $k$ is meaningless due to the missing hubs in each realization. Therefore one must employ for this comparison the ``cumulated probability'' method of Sect.\ \ref{ddd}. A typical result is shown in Fig.\ \ref{fig-knn-media}. Even with a small ensemble (e.g.\ 40 networks, in the figure) $\langle k_{nn} \rangle$ compares well with $k^{FR}_{nn}$.

The assortativity coefficient $r$ of the configuration model ensemble, which condensates the information on the correlations into a single number, {is for the examples given (BA2, $n=80$), $\left\langle r \right\rangle=-0.031$, with standard deviation $\sigma_r=0.028$.} The value of $r$ computed from the FR correlations is larger in absolute value: $r_{FR}\simeq -0.10$. The difference can probably be explained as due to the fluctuations.

\section{Maximally disassortative networks with scale-free exponent 3}
\label{max}

We have seen that finite BA networks are moderately disassortative, and not uncorrelated as frequently stated in the literature. A natural question arises: how relevant is this disassortativity? For scale-free networks with the same exponent ($\gamma=3$) what is the lowest possible value of $r$ attainable? And what is the aspect of networks with such minimum $r$, compared to BA networks? We recall that the $r$ coefficient of real-world {scale-free} disassortative networks is rarely more negative than $-0.15$, see \cite{bertotti2018bass}{, even if in general the $r$ coefficient of biological and technological networks can be smaller, especially for small size networks. For example, the protein-protein interaction network of H.\ pylori ($N=709$) has $r=-0.243$ and the protein-protein interaction network of C.\ elegans ($N=2386$) has $r=-0.183$. This fact has important consequences for the spectra of the networks \cite{jalan2015assortative}.} 

The configuration model offers a powerful tool for the exploration of such issues. In this section we shall show that it is possible to do an efficient rewiring of a network with the BA degree distribution which decreases the value of $r$, and consistently yields a minimum value which is approximately {three times} the value of the $r$ coefficient of the corresponding BA network. {For instance, for a maximum degree $n=80$ and $\beta=2$ we have $r\simeq -0.32$.} This makes sense for BA networks with $\beta \ge 2$, because for $\beta=1$ the giant component obtained from the configuration model is not complete.

With the pseudo-random wiring process described in Sect.\ \ref{wir} we obtain a network in the form of a list of $L$ links. We shall now perform on this network a rewiring process with the aim of increasing or decreasing its assortativity coefficient $r$ until respectively a maximum or minimum are reached. Each elementary rewiring step works as follows.

(1) Two links are chosen at random in the list. Suppose the first link is between nodes $a$ and $b$ and the second between nodes $c$ and $d$ ($a,b,c,d=1,\ldots,N$). Let the excess degrees of these nodes be respectively $A,B,C,D$.

(2) The links $[a,b]$ and $[c,d]$ are replaced by new links $[a,c]$ and $[b,d]$, provided $a\neq c$ and $b\neq d$ (in order to avoid the formation of loops). This rewiring step causes a change in the elements of the $e_{jk}$ matrix for which $j$ or $k$ coincide with $A,B,C$ or $D$. We recall that after the wiring process the $e_{jk}$ matrix is computed by counting in the node list, for each fixed couple of values $j,k$, how many links are between a node with degree $j$ and one with degree $k$, and dividing by $2L$; nodes with $j=k$ are counted twice. The variations after one rewiring step are
\begin{equation}
\Delta e_{AB}=-\frac{1}{L} \, ; \ \ \ \Delta e_{CD}=-\frac{1}{L} \, ; \ \ \
\Delta e_{AC}=\frac{1}{L} \, ; \ \ \ \Delta e_{BD}=\frac{1}{L} \, . \ \ \
\end{equation}

The variation of the Newman assortativity coefficient $r$ after one rewiring step is due to the presence in the definition of $r$ (eq.\ (\ref{asscoeffr}) in the Appendix) of the sum $\sum_{j,k=0}^{n-1}jke_{jk}$, which changes as
\begin{equation}
\Delta \sum_{j,k=1}^{n-1}jke_{jk}=2L^{-1}(-AB-CD+AC+BD) \equiv \Delta E \, .
\end{equation}
On the other hand, the sum $\sum_{j,k=0}^{n-1}jkq_jq_k$ does not change in the rewiring, because the distribution $q_k$ of the excess degrees does not change (it is a degree-preserving rewiring). For the same reason, the (positive) denominator in the definition of $r$ is also unchanged. The variation of $r$ is therefore
\begin{equation}
\Delta r=\frac{\Delta E}{\sum_{j,k=0}^{n-1}k^2q_k-\left(\sum_{j,k=0}^{n-1}kq_k\right)^2} \, .
\end{equation}
The variation is accepted when $r>0$, if we are looking for the maximum assortativity, or viceversa. The algorithm performs a large number of rewirings, for instance $10^5$ rewirings for a network with $\sim 10^3$ nodes; then $r$ is visualized and another $10^5$ rewirings are performed, and so on, until the value of $r$ stabilizes (this can be checked visually or through some automated criterium; the convergence is usually quick, and we shall discuss in further work the issue of possible local maxima and minima and how to exclude them).

\begin{figure}
\begin{center}
\includegraphics[width=8cm,height=5.5cm]{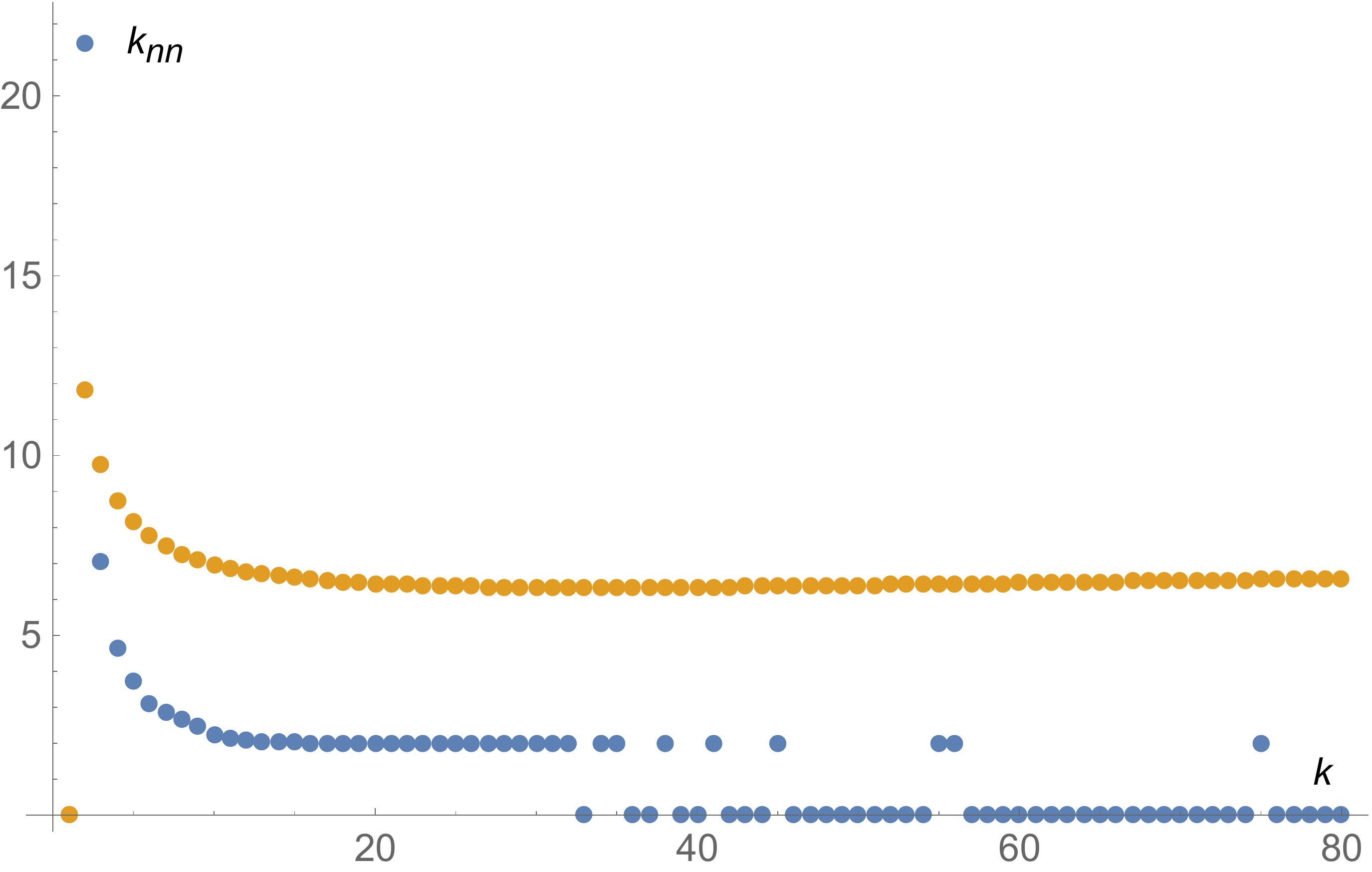}
\caption{Function $k_{nn}$ (dark point series) for a ``maximally disassortative'' network with 2500 nodes ($r\simeq -0.32$) obtained in the configuration model by a rewiring process which minimizes the assortativity coefficient $r$, starting from a BA2 degree distribution. The light point series represents, as in Fig.\ \ref{fig-knn-media}, the function $k_{nn}$ of an ideal BA2 ($r\simeq -0.10$). The hubs of the configuration model network are generated with the ``random hubs'' method. For the largest hubs of the maximally disassortative network one has $k_{nn}=2$ exactly, implying that all their nearest neighbours have degree 2. 
} 
\label{fig5}
\end{center}  
\end{figure}

In Fig.\ \ref{fig5} the function $k_{nn}$ which arises from one realization of a maximally disassortative network obtained with this method is compared with the corresponding $k_{nn}$ of a BA2 network.

The values of $r$ obtained will depend (for fixed $N$) on the degree distribution, therefore on the scale-free exponent $\gamma$ in the case of a pure power law, or on $\beta$ for a ``BA-like'' degree distribution 
$P(k)= 2 \beta(\beta+1)/[k(k+1)(k+2)]$. We use this degree distribution as a variation of the pure power law $\gamma=3$, in order to investigate the role of the details of the degree distribution at small $k$; 
we recall that these details influence the average degree and may have a strong impact, for instance, on the giant component of random networks in the configuration model \cite{newman2010networks}.

\section{Conclusion}
\label{conclusion}

The degree distribution, correlation functions and assortativity character are distinctive features of any network, and affect in an essential way the dynamics processes which take place on it. It is therefore desirable to develop methods and algorithms which generate networks where such characters are pre-assigned; this allows to study the resulting networks in detail and to simulate dynamical processes on them.

The configuration model \cite{newman2010networks,barabasi2016network} in its traditional form allows to generate uncorrelated networks with assigned degree distribution and has been widely investigated -- even though, for the scale-free case, defining the discretized degree distribution of the high-degree ``stubs'' in accordance with the integral criterium of Dorogovtsev-Mendes is not trivial (a point that we also fix in this paper, before addressing the correlations).

An improvement of the configuration model through a rewiring algorithm that generates an ensemble of networks with pre-assigned correlations has been proposed by Newman in his seminal paper on assortative mixing \cite{newman2003mixing}. In that paper Newman applied his rewiring algorithm to scale-free networks of the disassortative kind (and also of the assortative kind, in a small range of the $r$ coefficient). No further applications of this method have been published, to our knowledge; degree-preserving rewirings have been often used \cite{xulvi2004reshuffling,van2010influence,d2012robustness}, but not in connection with the correlation functions. Therefore the recent full computation of the correlations for BA networks \cite{fotouhi2013degree} offers the possibility of a new test of the Newman rewiring by comparison with the BA networks generated directly via the preferential attachment scheme.

In particular, we have tested numerically (and we plan to extend this work to other classes of correlation functions, besides those of BA networks): 
(a) the giant connected component of the networks obtained;
(b) the average of the function $k_{nn}$ in their ensemble;
(c) the fluctuations of $k_{nn}$ in the ensemble.

Furthermore, we have developed a new rewiring criterium which allows to obtain in an almost-deterministic way (i.e., with very small fluctuations in the resulting ensemble) networks having maximum or minimum values of the $r$ assortativity coefficient.

Finally, with this method it is also possible to focus on features related to specific components of the correlations. For instance, we observe that in the configuration model of BA1 networks the isolated couples are completely absent, thanks to the vanishing of the $e_{00}$ correlation; in a maximally disassortative network with BA2 degree distribution all the largest hubs are connected exclusively with nodes of degree 2, etc. These features have been obtained in trial networks with 2500 nodes, and therefore with a statistical significance of the order of 1 part in $10^3$. In future computations we plan to increase the accuracy and especially to address the case of assortative networks.

\section{\bf Abbreviations used}

BA: Barabasi-Albert

BA1, BA2, ...: Barabasi-Albert networks with $\beta=1,2,...$

UNC: Uncorrelated

ASS: Assortative

DIS: Disassortative 

\section{Declarations}

\bigskip

\noindent
{\bf Availability of data and materials:} - not applicable (no data and materials have been used or generated in this work).

\bigskip

\noindent
{\bf Competing interests} - None of the authors have any competing interests in the manuscript.

\bigskip

\noindent
{\bf Funding} - This work was supported by the Open Access Publishing Fund of the Free University of Bozen-Bolzano.

\bigskip

\noindent
{\bf Authors' contributions} - All authors contributed equally to this work.

\bigskip

\noindent
{\bf Acknowledgments} - Not applicable.

\section{Appendix}
\label{app}


We recall here some formulae which have been used throughout the paper and in the simulations.

Whereas $P(k)$ expresses the probability that a randomly chosen node of a network has degree $k$, 
the degree correlation $P(h | k)$ expresses the conditional probability that a node with degree $k$ is connected to one with degree $h$.

In particular, for a Barabasi-Albert network with parameter $\beta \ge 1$ corresponding to the number of parent nodes in the preferential attachment scheme,
the degree distribution is given \cite{barabasi2016network} by
\begin{equation}
P(k) =  \frac{2 \beta (\beta + 1)}{k (k+1) (k+2)} \, ,
\end{equation}
and the degree correlations are given by
\begin{equation}
	P(h | k) = \frac{\beta}{kh} \bigg( \frac{k+2}{h+1} - B^{2\beta+2}_{\beta+1} \, \frac{B^{k+h-2\beta}_{h-\beta}}{B^{k+h+2}_h} \bigg) \, ,
	\label{PhkBA}
\end{equation}
with $B^m_j$ denoting the binomial coefficient 
\begin{equation}
	B^m_j=\frac{m!}{j! (m-j)!} \, .
\end{equation}
The expressions in (\ref{PhkBA}) have been established by Fotouhi and Rabbat in \cite{fotouhi2013degree}
and have been employed by us, here and in \cite{bertotti2018bass}, after a suitable normalization,
due to the fact that we deal with network with a finite maximal degree.

\smallskip

As for the assortativity  coefficient $r$, its expression is given (see e.g. \cite{newman2002assortative} or \cite{bertotti2018bass}) by
\begin{equation}
	r = \frac{1}{\sigma^2_q} \, \sum_{k,h=0}^{n-1} kh (e_{kh}-q_kq_h) \, ,
	\label{asscoeffr}
\end{equation}
where $e_{kh}$ denotes the probability that a randomly chosen link connects nodes with excess degree $k$ and $h$ (the excess degree of a node being its total degree minus one),
and $q_k$ and $\sigma^2_q$ are given by
\begin{equation}
	q_k=\frac{(k+1)}{\sum_{j=1}^n j P(j)} \, P(k+1) 
\end{equation}
and
\begin{equation}
	\sigma^2_q = \sum_{k=1}^n k^2 q_k - \bigg(\sum_{k=1}^n k q_k \bigg)^2 \, .
\end{equation}

\bibliographystyle{unsrt}
\bibliography{jb3refsfornewman}

\end{document}